\documentclass[%
 reprint,
 amsmath,amssymb,
 aps,
]{revtex4-1}
\usepackage{verbatim}
\usepackage{graphicx}% Include figure files
\usepackage{dcolumn}% Align table columns on decimal point
\usepackage{bm,amsmath}% bold math
\begin{document}

\preprint{APS/123-QED}

\title{A new scheme based on the Hermite expansion to construct \\
lattice Boltzmann models associated with arbitrary specific heat ratio}% Force line breaks with \\

\author{Kainan Hu}
 \altaffiliation[Also at ]{University of Chinese Academy Sciences,Beijing ,China}%Lines break automatically or can be forced with \\
\author{Hongwu Zhang}%
 \email{Corresponding author : zhw@iet.cn}
\affiliation{%
Industrial Gas Turbine Laboratory, Institute of Engineering Thermophysics,\\
 Chinese Academy of Sciences,Beijing, China
}%

\author{Shaojuan Geng}
\affiliation{
Industrial Gas Turbine Laboratory, Institute of Engineering Thermophysics,\\
 Chinese Academy of Sciences, Beijing, China
}%

%\author{Yonghao Zhang}
%\affiliation{
% Mechanical and Aerospace Engineering,University of Strathclyde, Glasgow,UK
%}%

%\author{Jianping Meng}
%\affiliation{
% UK
%}%

\date{\today}% It is always \today, today,

\begin{abstract}

A new lattice Boltzmann scheme associated with flexible specific heat ratio is proposed. The new free degree is introduced via the internal energy associated with the internal structure. The evolution equation of the distribution function is reduced to two  evolution equations. One is connected to the density and velocity, the other is of the energy.
%Lattice Boltzmann models are derived  via the Hermite expansion. % from the local equilibrium function.
A two-dimensional lattice Boltzmann model and a three-dimensional  lattice Boltzmann model are derived via the Hermite expansion.
% of the equilibrium distribution function.
The two lattice Boltzmann models are applied to simulating the  shock tube of one dimension. Good agreement between the numerical results and the analytical solutions are obtained.

\begin{description}
\item[PACS numbers]
47.11.-j, 47.10.-g, 47.40.-x
\end{description}
\end{abstract}

\pacs{47.11.-j, 47.10.-g, 47.40.-x}% PACS, the Physics and Astronomy
\maketitle
%\tableofcontents

%\section{\label{sec:level1}First-level heading:\protect\\ The line
%break was forced \lowercase{via} \textbackslash\textbackslash}

\section{\label{sec:sec1}Introduction}
A lot of thermal lattice Boltzmann(LB) models  have been proposed in recent twenty years.
% All of the thermal LB models can be classified into two kinds.
% The first kind is based on the Hermite expansion and the second kind  has no relation  to the Hermite expansion.
The early thermal LB models are constructed by a try-error way. The discrete velocity set and the local equilibrium distribution function are determined by a set of constraints which makes sure the macroscopic equations match the thermohydrodynamic equations with certain accuracy\cite{Alexander1993Lattice,Qian1993Simulating,1994PhRvE..50.2776C}.
%This is an old-fashioned way to construct LB models.
Since 2006, a new method to construct LB models based on the Hermite expansion has been developed\cite{philippi2006continuous,philippi2015high,mattila2014high,shan2006kinetic,shan2010general,
shim2011thermal,shim2012uniform,shim2013multidimensional,shim2013univariate,ansumali2003minimal,chikatamarla2006entropy,
chikatamarla2009lattices}. The  LB models based on the Hermite expansion are more stable and it is convenient to construct LB models of any required level of accuracy.

Although much effort has been devoted to construct  thermal LB models and great development has been made, there are still  some problems to be resolved.
One of them is that the specific heat ratio $\gamma$
associated with these lattice models is fixed, in other words, the specific heat ratio $\gamma$ is not realistic.

To construct LB models associated with flexible specific heat ratio, several attempts  have been made\cite{shi2001finite,kataoka2004lattice,watari2007finite,tsutahara2008new}.
All these thermal LB models are constructed by the try-error way and besides the translational velocity of particle, the new introduced variables, such as the thermal energy or the rotational velocity of particle, are also discretized. So although the specific heat ratio is flexible, these LB model are much more complex than the ones associated with fixed specific heat ratio.
%So these thermal LB models are different from those ones with fixed specific heat ratio.

Another problem of the existing LB schemes associated with flexible specific heat ratio is that there are some drawbacks with the three-dimensional(3D) formation of these LB models. The 3D formation of the LB model proposed by \cite{kataoka2004lattice} is not stable. The LB scheme proposed by\cite{watari2007finite} can not  derive the Navier-Stokes  equations  in 3D via the Hermite expansion; only the Euler equations can  be derived.

%The general way construct LB models, such as the Hermite expansion, can not be applied to constructing LB models with arbitrary specific heat ratio except some special treatments are made.

This work propose a new LB scheme associated with arbitrary specific heat ratio. The evolution equation of the distribution function is reduced to two evolution equations. One of the reduced evolution equation is related to the translational velocity of particle and the other is connected with the energy distribution function. Unlike the LB schemes mentioned above, only the translational velocity of particle is discretized in  discrete velocity space.  The discrete particle velocity sets are same as the ones associated  with fixed specific heat ratio. The  discrete velocity set and the equilibrium distribution function are derived
via the Hermite quadrature and the Hermite expansion. The LB model proposed by this work is more stable than the ones constructed by the try-error way and it is easy to construct LB models of any required level of accuracy.

The 3D formation of the LB scheme proposed by this work is stable and the Navier-Stokes equations in 3D can be derived via the Hermite expansion.

The LB scheme proposed by this work is validated  by the  shock tube problem of one dimension. A 2D lattice model and  a 3D lattice model are  employed to simulate the shock tube flow. The results of simulation agree with the analytical solutions very well.
\vspace{1.8cm}

\section{Two reduced Boltzmann BGK equations}
The origin that the specific heat ratio is fixed is that  gases are supposed to be monatomic, so there is only the translational free degree. To describe  diatomic gases or polyatomic gases, a  parameter $I$ connected with the internal energy
%which takes into the degrees of freedom
should be introduced into the distribution function\cite{lifschitz1983physical}.

The polyatomic distribution function $f(\bm{\xi},I,\bm{x},t)$ is the probability density with the particle velocity $\bm{\xi}$ and internal energy $\epsilon(I) = I^{2/\delta}$ at point $\bm{x}$, time $t$.
The density $\rho$, flow velocity $\bm{u}$ and total energy $E$ are defined as % heat flux $\bm{q}$ and stress tensor $\bm{\tau}$ are defined by the moment of distribution function i.e.
\begin{subequations}
\begin{align}
\rho = &\int_{R^3}\!\int_{R^+} f(\bm{\xi},I) d \bm{\xi} d I ,\label{Eq:FeqRhoXi}\\
\rho \bm{u}= &\int_{R^3}\!\int_{R^+} f(\bm{\xi},I) \bm{\xi} d \bm{\xi} d I ,\label{Eq:FeqUXi}\\
\rho E =& \int_{R^3}\!\int_{R^+} f(\bm{\xi},I) (\frac{1}{2}\xi^2 + I^{2/\delta}) d \bm{\xi} dI .\label{Eq:FeqEXi}
\end{align}
\end{subequations}
where $E\!=\!\displaystyle \rho{\frac{1}{2} u^2 + e_{tr} + e_{int}}$ is the specific total energy, $e_{tr} = \displaystyle\frac{D}{2} R_g T$ is the specific translational energy, $D$ is the dimension, $T$ is the absolute temperature, $R_g$ is the universal gas constant, $e_{int} = \displaystyle\frac{\delta}{2} R_g T$ is the energy associated with the internal structure. The specific internal energy $e$ is defined as
%where $E\!=\!\displaystyle \rho{\frac{1}{2} u^2 + e_{tr} + e_{int}}$ is the specific total energy, $e_{tr}$ is the specific translational energy,  $e_{int} $ is the energy associated with the internal structure.  The internal energy $e$ is defined as
$e =  e_{tr} + e_{int}$ and
\begin{subequations}
\begin{align}
 e_{tr} =  & \frac{1}{\rho} \int_{R^3}\!\int_{R^+} \frac{1}{2}(\bm{\xi}-\bm{u})^2 f d \bm{\xi} d I ,\\
 e_{int}& = \frac{1}{\rho} \int_{R^3}\!\int_{R^+} I^{2/\delta} f d \bm{\xi} d I.
\end{align}
\end{subequations}

Here, the specific heat ratio $\gamma$ is defined as
\begin{equation}
\gamma = \frac{\delta + D +2}{\delta + D}.
\end{equation}

The equilibrium distribution function $f^{eq}$ can be expressed\cite{lifschitz1983physical}
\begin{align}
f^{eq}(\bm{\xi},I,\bm{x},t) = & \Lambda_{\delta} \rho \frac{1}{(2\pi R_g T)^{\frac{D}{2}}}\frac{1}{( R_g T)^{\frac{\delta}{2}}}\notag\\ &\times \exp{\Big[-\frac{(\bm{\xi}-\bm{u})^2}{2 R_g T} - \frac{I^{\frac{2}{\delta}}}{R_g T} \Big]},
\end{align}
where $\Lambda_{\delta}^{-1} = \int \exp( -I^{\frac{2}{\delta}}) d I  $ is a constant,
$\rho$ is the density.

%The variables above is dimensional. For convenience, we define the dimensionless variables.
The dimensionless formation of the  equilibrium distribution function is
\begin{align}\label{Eq:DDF}
\tilde f^{eq}(\tilde{\bm{\xi}},\tilde{I},\tilde{\bm{x}},\tilde{t}) =& \tilde \Lambda_\delta \frac{\tilde \rho}{(2 \pi \tilde \theta)^\frac{D}{2}}
\frac{1}{\tilde \theta^{\frac{\delta}{2}}} \notag\\
&\times \exp  \Big(-\frac{|\tilde {\bm{\xi}}- \tilde {\bm{u}}|^2}{2 \tilde{\theta}}\Big) \exp \Big( -\frac{\tilde I^{\frac{2}{\delta}}}{\tilde \theta }\Big),
\end{align}
where
\[ \begin{array}{lll}
&\tilde f^{eq}\!=\! f^{eq}\displaystyle\theta_0^{\frac{N}{2}} \displaystyle\theta_0^{ \frac{\delta}{2}}/\rho_0,
                                        &\tilde \Lambda_{\delta}^{-1}\!=\!\int \exp( -\tilde I^{\frac{2}{\delta}}) d \tilde I,\\
&\tilde I\!=\!I/\theta_0^{\frac{\delta}{2}},     & \tilde \rho\!=\!\rho / \rho_0,  \\              &\tilde{\bm{\xi}}\!=\!\bm{\xi}/\sqrt{\theta_0},  &\tilde{\bm{u}}\!=\!\bm{u}/\sqrt{\theta_0}, \\
&\tilde T = \tilde \theta\!=\!\theta / \theta_0, &\theta\!=\!R_g T, \\
&\tilde{\bm{x}}\!=\!\bm{x}/(\bm{\xi}_0 t_0),     &\tilde{\bm{\xi}}\!=\!\bm{\xi}/\bm{\xi}_0,\\
&\tilde t\!=\!t/t_0,                             &t_0\!=\!L_0/\sqrt{\theta_0}, \\
&\theta_0\!=\!R_g T_0,\\
\end{array}\]
$L_0$ is the characteristics length,
$T_0$ is the characteristics temperature,
$\rho_0$ is the characteristics density,
$t_0$ is the characteristics time.

The other dimensionless  variables are defined as
\[ \begin{array}{lll}
&\tilde f\!=\!f\displaystyle\theta_0^{\frac{N}{2}} \displaystyle\theta_0^{ \frac{\delta}{2}}/\rho_0,
                                                    &\tilde E\!=\!E/\theta_0, \\
&\tilde e\!=\!e/\theta_0,                           &\tilde e_{tr}\!=\!e_{tr}/\theta_0,    \\
&\tilde e_{int}\!=\!e_{int}/\theta_0,               &\tilde p\!=\!p/(\rho_0\theta_0).      \\
\end{array}\]

In the following part, the tildes are omitted and
the dimensionless distribution function can be expressed
\begin{align}\label{Eq:DDFDimensionless}
 f^{eq}({\bm{\xi}},{I},{\bm{x}},&{t}) =  \Lambda_\delta \frac{ \rho}{(2 \pi  \theta)^\frac{D}{2}}
\frac{1}{ \theta^{\frac{\delta}{2}}}\notag\\
 &\times \exp  \Big(-\frac{| {\bm{\xi}}-  {\bm{u}}|^2}{2 {\theta}}\Big) \exp \Big( -\frac{ I^{\frac{2}{\delta}}}{ \theta }\Big).
\end{align}

Accordingly, the dimensionless state equation is $p = \rho T$, where $p$ is the pressure. The dimensionless internal energy $e$, translational energy $e_{tr}$ and the energy connected with the structure $e_{int}$ are defined as $e = \displaystyle\frac{D+\delta}{2}T$, $e_{tr} = \displaystyle\frac{D}{2}T$, $e_{int}=\displaystyle\frac{\delta}{2}T$.
The dimensionless evolution equation of the distribution function $f(\bm{\xi},I,\bm{x},t)$ can be expressed as
\begin{equation}\label{Eq:EvolF}
\frac{\partial f}{\partial t} + \bm{\xi} \cdot \nabla f = -\frac{1}{\tau}(f-f^{eq}),
\end{equation}
where $\tau$ is the relaxation time.

For the sake of obtaining correct specific heat ratio $\gamma$, the parameter $I$($\epsilon$)  is introduced. This is  similar with the existing LB models associated with flexible specific heat ratio. In those LB models, a new parameter such as the rotational velocity or the rotational energy is introduced. However, unlike the existing schemes\cite{kataoka2004lattice}\cite{watari2007finite}, the LB scheme given by this work does not dicretize the new parameter on lattices. Instead, only the translational  velocity of particle is discretized. In the LB scheme proposed by this work, two reduced distribution function, i.e. the distribution function of mass $g(\bm{\xi})$  and that of  energy $h(\bm{\xi})$ are introduced. This idea  is first proposed by C.K.Chu\cite{chu1965kinetic}\cite{chu1965kinetic2}. His  aim is to save computational resource. When the idea is applied to the lattice Boltzmann method, there is another advantage that it is not necessary to discretize  $I(\epsilon)$ on lattices. The lattices employed in the scheme proposed by this work is same as the ones associated with fixed specific heat ratio. These lattice models is easier to construct than the ones in which both the translational velocity of particle and the new introduced parameter need to be discretized. %The scheme proposed by this work is convenient to apply to the three dimension flow.

The  procedure of obtaining the reduced distribution functions  is as following. We begin with the definitions of the two reduced distribution functions
\begin{subequations}
\begin{align}
g(\bm{\xi}) = &\int_{R^{+}} f(\bm{\xi},I) d I, \\
h(\bm{\xi}) = &\int_{R^{+}} I^{\frac{2}{\delta}} f(\bm{\xi},I) d I.
\end{align}
\end{subequations}

The macroscopic variable are defined by the moments of the distribution function of mass $g(\bm{\xi})$ and the distribution function of the  energy $h(\bm{\xi})$
\begin{subequations}
\begin{align}
\rho = &\int_{R^{3}} g(\bm{\xi}) d \bm{\xi},\\
\rho \bm{u} =& \int_{R^{3}} \bm{\xi} g(\bm{\xi}) d \bm{\xi},\\
\rho E = &\int_{R^{3}} [ \frac{1}{2}\xi^2 g(\bm{\xi}) + h(\bm{\xi}) ] d\bm{\xi}.
\end{align}
\end{subequations}

The associated Maxwell-Boltzmann distribution is defined as
\begin{align}
F^{eq}(\bm{\xi})
= &\int_{R^+} f^{eq}(\bm{\xi},I) d I \notag\\
= &\int_{R^+} \Lambda_\delta \frac{ \rho}{(2 \pi  \theta)^\frac{D}{2}}\frac{1}{ \theta^{\frac{\delta}{2}}} \notag\\
  &\times \exp  \Big(-\frac{| {\bm{\xi}}-  {\bm{u}}|^2}{2 {\theta}}\Big)
    \exp \Big( -\frac{ I^{\frac{2}{\delta}}}{ \theta }\Big) d I \notag\\
= &\rho \frac{1}{(2\pi \theta)^{\frac{D}{2}}}  \exp{\Big[-\frac{(\bm{\xi}-\bm{u})^2}{2 \theta} \Big]}.
\end{align}

Integrating Formula(\ref{Eq:EvolF}) on $I$, we obtain the evolution equation of $g(\bm{\xi})$
\begin{align}\label{Eq:ReductionG}
\frac{\partial }{\partial t}g(\bm{\xi}) + \bm{\xi}\cdot\nabla g(\bm{\xi}) = -\frac{1}{\tau}[g(\bm{\xi}) - F^{eq}(\bm{\xi})].
\end{align}

Integrating Formula$(\ref{Eq:EvolF}) \times I^{\frac{2}{\delta}}$ on $I$, we obtain the evolution equation of $h(\bm{\xi})$
\begin{align}\label{Eq:ReductionH}
\frac{\partial }{\partial t}h(\bm{\xi}) + \bm{\xi}\cdot\nabla h(\bm{\xi}) = -\frac{1}{\tau}[h(\bm{\xi}) - \frac{\delta}{2} T F^{eq}(\bm{\xi})].
\end{align}

Discretizing Formula(\ref{Eq:ReductionG}) and (\ref{Eq:ReductionH}) in  discrete velocity  space, we obtain two discrete reduced evolution equations
\begin{subequations}
\begin{align}
\frac{\partial }{\partial t}g_i + \bm{\xi}_i\cdot\nabla g_i =& -\frac{1}{\tau}(g_i - F_i^{eq}),\label{Eq:DisReductionG}\\
\frac{\partial }{\partial t}h_i + \bm{\xi}_i\cdot\nabla h_i =& -\frac{1}{\tau}(h_i - \frac{\delta}{2} T F_i^{eq}).\label{Eq:DisReductionH}
\end{align}
\end{subequations}
%From Formula(\ref{Eq:DisReductionG}) and (\ref{Eq:DisReductionH}), the Navier-Stokes equations associated with flexible specific heat ratio can be derived via the Chapman-Enskog expansion.

In discrete velocity space, the density $\rho$, the macroscopic velocity $\bm{u}$ and the  specific total energy $E$  are defined as
\begin{subequations}
\begin{align}
\rho =& \sum_i g_i, \\
\rho \bm{u} =& \sum_i g_i \bm{\xi}_i, \\
\rho E =& \sum_i g_i \frac{1}{2}\xi^2_i + \sum_i h_i.\label{Eq:TransEnergy}
\end{align}
\end{subequations}
%where $ E\!=\!\displaystyle \frac{1}{2}u^2 + e$,
%$e = e_{tr} + e_{int}$ is the specific internal energy, $e_{tr}\!=\!\displaystyle \frac{D}{2}T$ is the translational energy,
%$e_{int}\!=\!\displaystyle \frac{\delta}{2} T$ is the specific internal energy connected with the  internal structure.

We also have the relationships
\begin{subequations}
\begin{align}
\rho (\frac{1}{2}u^2 + e_{tr}) =& \sum_i g_i \frac{1}{2}\xi^2_i, \\
%\frac{1}{2}\rho e_{int} =& \sum_i h_i
\rho e_{int} =& \sum_i h_i.
\end{align}
\end{subequations}

\section{Calculation procedure}

 We employ the first order difference to discretize the reduced evolution of $g_i$, i.e. Formula(\ref{Eq:DisReductionG}), in time
\begin{equation}\label{Eq:DiscreteReductionG}
g_i(\bm{x},t + \Delta t) = g_i(\bm{x},t) - \Delta t \bm{\xi}_i \cdot \nabla g_i -  \frac{\Delta t}{\tau}[g_i(\bm{x},t) -F_i^{eq}(\bm{x},t)],
\end{equation}
where $\Delta t$ is the time step, $\Delta x$ is the grid step.
The convection term along the coordinate $x$ is performed by the third order upwind scheme
\[
\begin{split}
\xi_{ix} \frac{\partial g_{i}}{\partial x} = &\frac{1}{2}\frac{\xi_{ix}+|\xi_{ix}|}{6\Delta x}[g_i(x\!-\!2\Delta x)- 6g_i(x\!-\!\Delta x)\\
&+ 3g_j(x) + 2g_j(x\!+\!\Delta x)] \\
&+\frac{1}{2}\frac{\xi_{ix}-|\xi_{ix}|}{6\Delta x}[-g_i(x\!+\!2\Delta x) + 6g_i(x\!+\!\Delta x)\\
&- 3g_i(x) - 2g_i(x\!-\!\Delta x)].
\end{split}
\]
The discretized  convection terms along the $y$ and $z$ coordinate are similar.

In a similar way,  the discretized form of the discrete reduced evolution equation of $h_i$, i.e. Formula(\ref{Eq:DisReductionH}) can be obtained,
\begin{align}\label{Eq:DiscreteReductionH}
h_i(\bm{x},t + \Delta t) &= h_i(\bm{x},t) - \Delta t \bm{\xi}_i \cdot \nabla h_i\notag\\
& -  \frac{\Delta t}{\tau}[h_i(\bm{x},t) -\frac{\delta}{2}T F_i^{eq}(\bm{x},t)].
\end{align}
The convection term is same as that of $g_i$.

\section{\label{sec:sec3}LB models of 2D and 3D }
Employing  the Hermite quadrature, which has been intensively discussed in \cite{philippi2006continuous,philippi2015high,mattila2014high},\cite{shan2006kinetic,shan2010general}, \cite{shim2013multidimensional,shim2013univariate},
we  construct a 2D LB model, i.e. D2Q37, and a 3D LB model, i.e. D3Q105. Both models are of  fourth-order accuracy. The discrete particle velocity sets and the weights $\omega_i$ of D2Q37 and D3Q105 are showed in Table(\ref{tab:D2Q37}) and Table(\ref{tab:D3Q107}) respectively. The discrete associated Maxwell-Boltzmann distribution  of fourth-order accuracy is
\begin{equation}
F^{eq}_i(\bm{\xi}) = \omega_i \rho \sum_{k=0}^{4}\frac{1}{k!}\bm{a}^{(k)}\cdot\bm{H}^{(k)},
\end{equation}
where
\begin{align}
 \bm{a}^{(0)}\cdot\bm{H}^{(0)}= & 1,\notag\\
 \bm{a}^{(1)}\cdot\bm{H}^{(1)}= & \bm{\xi} \cdot \bm{u},\notag\\
 \bm{a}^{(2)}\cdot\bm{H}^{(2)}= & (\bm{\xi} \cdot \bm{u})^2 + (\theta -1) (\theta^2 -D) - u^2,\notag\\
 \bm{a}^{(3)}\cdot\bm{H}^{(3)}= & (\bm{\xi} \cdot \bm{u})[(\bm{\xi} \cdot \bm{u})^2 - 3u^2 \notag\\
                                & + 3(\theta -1)( u^2 - D -2)],\notag\\
 \bm{a}^{(4)}\cdot\bm{H}^{(4)}= & (\bm{\xi} \cdot \bm{u})^4 - 6(\bm{\xi} \cdot \bm{u})^2 u^2 + 3u^4 \notag\\
                                & + 6(\theta -1)[(\bm{\xi} \cdot \bm{u})^2(u^2 - D - 4) \notag\\
                                & + (D + 2 - u^2 )\xi^2] \notag\\
                                & + 3(\theta -1)^2[u^4 - 2(D+2)u^2+D(D+2)],\notag
\end{align}
and D is the dimension.

\begin{table}[!htbp]
\caption{  Discrete velocities and weights  of D2Q37.
Perm denotes permutation and $k$ denotes the number of
 discrete velocities included in each group.Scaling factor is $r =\! 1.19697977$.}
 \begin{ruledtabular}
\begin{tabular}{ p{30pt} p{60pt}  p{80pt}   }

 $k$        & $\bm{\xi}_i$   & $\omega_i$      \\
\hline
  $1 $ & $(0,0)$       & $2.03916918e\!-\!1$      \\
  $4 $ & $Perm(r,0)$   & $1.27544846e\!-\!1$      \\
  $4 $ & $Perm(r,r)$   & $4.37537182e\!-\!2$      \\
  $4 $ & $Perm(2r,0)$  & $8.13659044e\!-\!3$      \\
  $4 $ & $Perm(2r,r)$  & $9.40079914e\!-\!3$      \\
  $4 $ & $Perm(3r,0)$  & $6.95051049e\!-\!4$      \\
  $4 $ & $Perm(3r,r)$  & $3.04298494e\!-\!5$      \\
  $4 $ & $Perm(3r,3r)$ & $2.81093762e\!-\!5$      \\
\end{tabular}
\end{ruledtabular}
\label{tab:D2Q37}
\end{table}

\begin{table}[!htbp]
\caption{  Discrete velocities and weights  of D3Q107.
Perm denotes permutation and $k$ denotes the number of
 discrete velocities included in each group.Scaling factor is $r\!=\! 1.19697977$.}
\begin{ruledtabular}
\begin{tabular}{ p{30pt} p{60pt}  p{80pt}   }

 $k$        & $\bm{\xi}_i$   & $\omega_i$      \\
\hline
  $1  $ & $(0,0,0)$          & $1.603003550e\!-\!1$      \\
  $6  $ & $Perm(r,0,0)$      & $2.390868500e\!-\!2$      \\
  $12 $ & $Perm(r,r,0)$      & $4.169870315e\!-\!2$      \\
  $8  $ & $Perm(r,r,r)$      & $5.166346780e\!-\!3$      \\
  $6  $ & $Perm(2r,0,0)$     & $1.227117147e\!-\!2$      \\
  $24 $ & $Perm(2r,r,r)$     & $2.676524495e\!-\!3$      \\
  $12 $ & $Perm(2r,2r,0)$    & $9.685223338e\!-\!4$      \\
  $8  $ & $Perm(2r,2r,2r)$   & $2.170763000e\!-\!5$      \\
  $6  $ & $Perm(3r,0,0)$     & $2.453010230e\!-\!4$      \\
  $24 $ & $Perm(3r,r,r)$     & $1.417071270e\!-\!4$      \\
\end{tabular}
\end{ruledtabular}
\label{tab:D3Q107}
\end{table}

%It should be noticed that the LB model given above is different from the LB model given by\cite{philippi2006continuous}.

\section{Numerical validation}
To validate the scheme proposed by this work, we apply the scheme to the  shock tube problem of one dimension\cite{SOD19781}. %This is a one dimension case
 A 2D LB model, i.e. D2Q37, and a 3D LB model, i.e. D3Q107,  are employed  to solve this problem.

The initial condition is given by
\begin{align}
&(\rho_L,T_L, u_{Lx})\!=\!(4,1,0)\\
&(\rho_R,T_R, u_{Rx})\!=\!(1,1,0)
\end{align}
where the subscript $L$ indicates the left side of the shock tube and  $R$ indicates the right side of the shock tube.
$u_x$ is the macroscopic velocity along the $x$ coordinate.

We set the specific heat ratio as $\gamma  \!=\!1.4$ and the relaxation time as $\tau\!=\!2/3$.
All of these macroscopic variables are dimensionless.

In the case of 2D, the grid is $X \times Y \!=\!  1000 \times 16 $. The parameter $\delta$ is $3$. The periodic boundary condition is employed for the up and down boundaries and the open boundary condition is employed for the left and right boundaries.

In the case of 3D, the grid is  $X \times Y \times Z \!=\!  1000 \times 16 \times 16$, and the $\delta = 2$.
The open boundary condition is employed for the left and right boundaries and the  periodic boundary condition is employed for the others.

\begin{figure*}[!hbt]
\includegraphics[width=16cm]{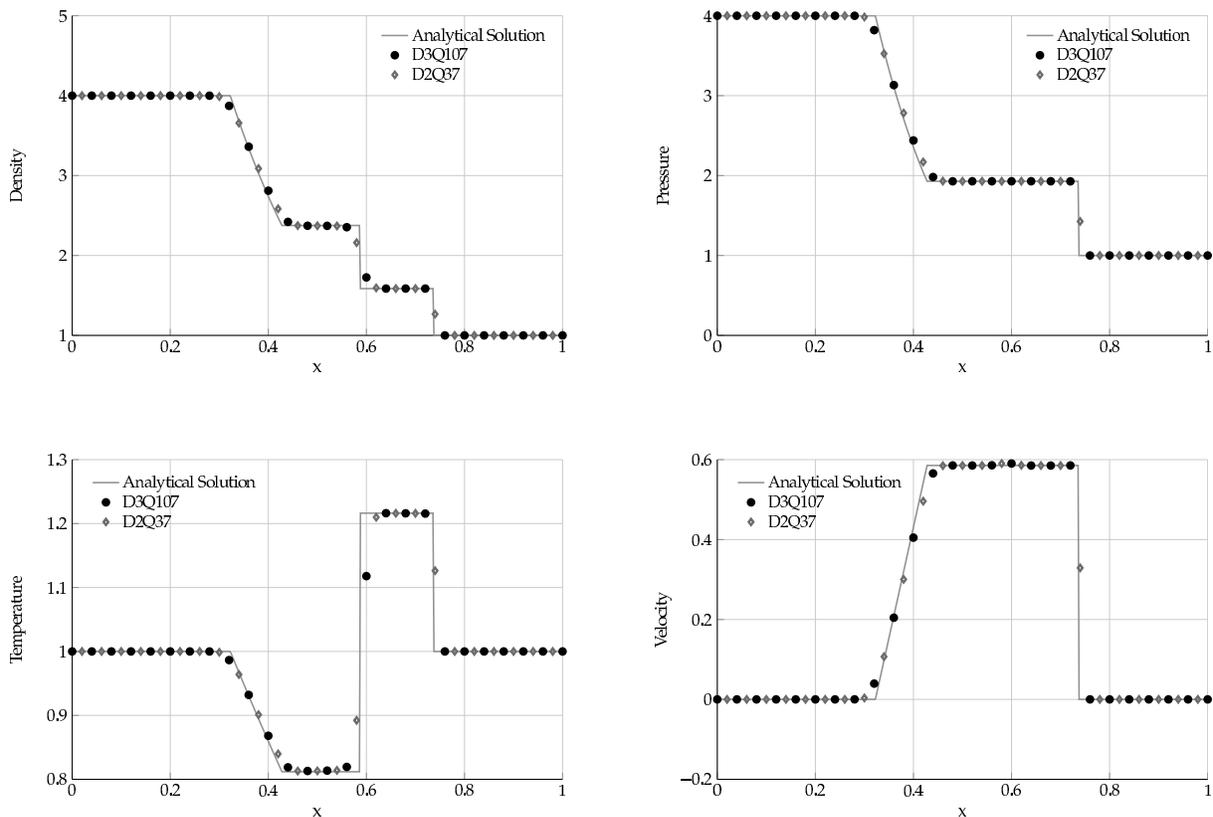}% Here is how to import EPS art
\caption{The simulation results of D2Q37, D3Q107, and the analytical resolutions  at the time $t\!=\!0.1504$. The specific heat ratio is $\gamma\! = \! 1.4$ and the parameter $\delta$ is $3$ in the case of 2D and $2$ in the case of 3D. The relaxation time is $\tau\!=\! 2/3$. The initial condition of the left side is ${\rho \!=\! 4,T\!=\!1,\bm{u}\!=\!0}$ and that of the right side is ${\rho \!=\! 1,T\!=\!1,\bm{u}\!=\!0}$.}
\label{Fig:HeatRatio2D}
\end{figure*}

Fig(\ref{Fig:HeatRatio2D})   show  the simulation results of 2D and 3D  at $step$ $180$, i.e. at the time  $$t = \displaystyle{\frac{step}{ X \times r}}= \displaystyle{\frac{180}{1000\times 1.1969797752}} = 0.1504 .$$
%The black lines show the simulation results and the gray lines show the analytical resolutions.
It  can be seen from Fig(\ref{Fig:HeatRatio2D}) that the simulation results agree with the analytical resolutions very well.

\begin{comment}
\begin{figure*}[!hbt]
\includegraphics[width=16cm]{D3Q107}% Here is how to import EPS art
\caption{The black lines are the simulation results and the gray lines are the analytical resolutions. These are the results at the 180th step, i.e. at the time $t\!=\!0.1504$. $\delta\!=\!2$. The specific heat ratio is $\gamma\! = \! 1.4$ and the relaxation time is $\tau\!=\! 2/3$. The initial condition of the left tube is ${\rho \!=\! 4,T\!=\!1,\bm{u}\!=\!0}$ and that of the right tube is ${\rho \!=\! 1,T\!=\!1,\bm{u}\!=\!0}$.}
\label{Fig:HeatRatio3D}
\end{figure*}
\end{comment}
\section{Conclusion}

A new LB scheme associated with flexible specific heat ratio is proposed. A 2D LB model and a 3D LB model are derived via the Hermite expansion. The evolution equation of the distribution function is reduced to two evolution equations. The specific heat ratio obtained by the scheme  is adjustable. The  shock tube is employed to validate the proposed LB scheme.
The simulation results agree with the analytical resolutions very well.

The LB models employed in the proposed scheme are derived via the Hermite quadrature, so the proposed scheme is more stable than the ones derived via the early method, which construct LB models in a try-error way. It is easy to construct LB models of any required level  of accuracy.

For the proposed LB scheme, only the translational velocity is discretized and no other variables need to be discretized. So the LB scheme  proposed by this work is much simpler than the existing schemes and much easier to be implemented.% associated with flexible specific heat ratio.

\nocite{*}

\bibliography{HeatRatio}% Produces the bibliography via BibTeX.

%\bibliography{GAMMA}
\end{document}